\documentclass[aps,prc,amsmath,amssymb,preprint,floats,nofootinbib,superscriptaddress,color,groupedaddres]{revtex4}
\usepackage[dvips]{graphicx}
\usepackage[dvips]{color}
\usepackage{dcolumn}
\usepackage{bm}

\newcommand{\bea}{\begin{eqnarray}}
\newcommand{\eea}{\end{eqnarray}}
\newcommand{\be}{\begin{equation}}
\newcommand{\ee}{\end{equation}}
\newcommand{\ba}{\begin{eqnarray}}
\newcommand{\ea}{\end{eqnarray}}

\begin{document}

\title{Scaling Function and Nucleon Momentum Distribution}

\author{J.A. Caballero}
\affiliation{Departamento de F\'{\i}sica At\'omica, Molecular y Nuclear, Universidad de Sevilla, 41080 Sevilla, SPAIN}
\author{M.~B.~Barbaro}

\affiliation{Dipartimento di Fisica Teorica, Universit\`{a} di
Torino and INFN,\\
Sezione di Torino, Via P. Giuria 1, 10125 Torino, Italy}
\author{A.~N.~Antonov}

\affiliation{Institute for Nuclear Research and Nuclear Energy, Bulgarian Academy of Sciences, Sofia 1784, Bulgaria}

\author{M.~V.~Ivanov}
\affiliation{Institute for Nuclear Research and Nuclear Energy, Bulgarian Academy of Sciences, Sofia 1784, Bulgaria}

\author{T.~W.~Donnelly}
\affiliation{Center for Theoretical Physics, Laboratory for Nuclear Science and Department of Physics,\\
Massachusetts Institute of Technology, Cambridge, MA 02139, USA}

\date{\today}


\begin{abstract}
  Scaling studies of
  inclusive quasielastic electron scattering reactions have been used in the past
  as a basic tool to obtain information on the nucleon momentum distribution in nuclei.
  However, the connection between the scaling function, extracted from the analysis
  of cross section data, and the spectral function only exists assuming
  very restricted approximations. We revisit the basic expressions
  involved in scaling studies and how they can be linked to the nucleon momentum
  distribution. In particular, the analysis applied in the past to
  the so-called scaling region, {\it i.e.,} negative values of
  the scaling variable $y$, is extended here to positive $y$,
  since a ``universal'' superscaling function has been extracted from
  the analysis of the separated longitudinal data. This leads to results that
  clearly differ from the ones based solely on the negative-$y$ scaling
  region, providing new information on how the energy and momentum are distributed in the
  spectral function.

\end{abstract}


\maketitle

\section{Introduction: basic aspects of scaling}\label{sect1}

Scaling studies of inclusive quasielastic (QE) electron-nucleus
scattering have largely been considered to provide a powerful tool
for extracting the momentum distribution of nucleons inside
nuclei~\cite{Day:1990mf,Ciofi87,Ciofi89,Ciofi91,Ciofi92,Ciofi96,Ciofi99}.
Such analyses have been applied to few-body systems, complex nuclei
and nuclear matter with an important effort devoted to estimating
binding corrections, and in particular, the high-momentum components
of the nucleon momentum distribution which are governed by Short
Range Correlations (SRC)~\cite{Ji89,Ciofi09}. However, caution
should be borne in mind for the conclusions reached, since a close
relationship between the momentum distribution and the scaling
function only emerges after some approximations are made. These are
linked not only to the general description of the electron
scattering reaction mechanism, but also to the integration limits
involved and the behavior of the spectral
function~\cite{Day:1990mf}.

The phenomenon of $y$-scaling emerges from the analysis of QE
$(e,e')$ reactions. The scaling function, defined as the QE $(e,e')$
differential cross section divided by an appropriate factor involving the
single-nucleon cross section~\cite{scaling88,Day:1990mf,DS199,DS299},
is shown to depend only on a
single variable, $y$, given as a particular combination of the two
independent variables in the process, namely the energy and momentum
transfers, $\omega$ and $q$. In the QE domain and for values of
$\omega$ and $q$ large enough, the basic mechanism in $(e,e')$
reactions on nuclei corresponds to elastic scattering from
individual nucleons in the nuclear medium with ``quasi-free''
ejection of a nucleon from the nuclear system. This implies that the
inclusive $(e,e')$ cross section is mainly constructed from the
exclusive $(e,e'N)$ process, including the contribution of all
nucleons in the target and integrating over all (unobserved) ejected
nucleon variables. In other words, QE scattering off a nucleus is
simply described as an incoherent sum of single-nucleon scattering
processes. This approach, which constitutes the basis of the Impulse
Approximation (IA), although being an over-simplified description of
$(e,e')$ reactions, has demonstrated its validity under appropriate
kinematic conditions. Mechanisms beyond the IA (correlations, Meson
Exchange Currents (MEC), re-scattering processes, {\it etc.}) may
play a significant role in electron scattering, and hence may lead
to non-negligible scaling violations.

The IA provides an intuitive explanation on how the scaling behavior
emerges from the analysis of data. In this case the QE $(e,e')$
cross section is given by \be
\left[\frac{d\sigma}{d\epsilon'd\Omega'}\right]_{(e,e')}=
\sum_{i=1}^{A}\int\!\!\!\int_{\Sigma(\omega,q)}p\,dp\,d{\cal E}\int
d\phi_{N_i}\left(
\frac{E_{N_i}}{qp_{N_i}^2}\right)\left[\frac{d\sigma}{d\epsilon'd\Omega'dp_{N_i}d\Omega_{N_i}}
\right]_{(e,e'N_i)} \, , \label{IA} \ee where the sum extends to all
nucleons in the target and $\{\epsilon',\Omega'\}$ refer to the
scattered electron variables. The integration over the ejected
(unobserved) nucleon variables $\{p_{N_i},E_{N_i},\Omega_{N_i}\}$
has been expressed in terms of the residual nucleus' excitation
energy ${\cal E}$ and the missing momentum $p$. The significance of
these variables as well as the kinematically allowed integration
region denoted by $\Sigma(\omega,q)$ will be discussed in detail in
next section.

Within the IA the evaluation of $(e,e'N_i)$ cross sections for both
proton and neutron knock-out determines the inclusive QE cross
section. The study of exclusive $(e,e'N)$ reactions has been
presented in previous
work~\cite{Udias,highp,Udias3,Fru84,Bof96,Kel96} focusing on
different aspects of the problem: Final-State Interactions (FSI),
relativity, correlations, {\it etc.} Although such ingredients have
been proven to be essential in order to fit experimental $(e,e'N)$
cross sections, in what follows we restrict our attention to the
Plane-Wave Impulse Approximation (PWIA), where the knocked-out
nucleon has no interaction with the residual nucleus. Being the
simplest approach to $(e,e'N)$ processes, PWIA retains important
relativistic effects that are essential in describing reactions at
high $q$ and $\omega$. Moreover, the $(e,e'N)$ differential cross
section in PWIA factorizes in two basic terms: the electron-nucleon
cross section for a moving, off-shell nucleon and the spectral
function that gives the combined probability to find a nucleon of
certain momentum and energy in the nucleus~\cite{Fru84,Bof96,Kel96}.
In general we can write \be
\left[\frac{d\sigma}{d\epsilon'd\Omega'dp_Nd\Omega_N}
\right]_{(e,e'N)}^{PWIA}= K\sigma^{eN}(q,\omega;p,{\cal
E},\phi_N)S(p,{\cal E}) \label{PWIA} \ee with $K$ a kinematical
factor~\cite{Ras89} and where $p$ is the missing momentum and ${\cal
E}$ the excitation energy, essentially the missing energy minus the
separation energy. It is important to point out that the
factorization property shown in Eq.~(\ref{PWIA}) no longer persists
if dynamical relativistic effects in the bound nucleons are
incorporated, {\it i.e.,} effects from the lower components in the
relativistic wave functions, even in the plane-wave
limit~\cite{Caballero-NPA,Cris}. Note that both the $eN$ cross
section and the spectral function depend on the two integration
variables in Eq.~(\ref{IA}), $p$ and ${\cal E}$. In order to show
how the scaling function emerges from PWIA, further assumptions are
needed. First the spectral function is assumed to be isospin
independent, and second $\sigma^{eN}$ is assumed to have a very mild
dependence on the missing momentum and excitation energy, which is
supported by the most commonly used off-shell cross
sections~\cite{Day:1990mf}. Hence the $eN$ cross section can be
evaluated at fixed values of $p$ and ${\cal E}$: typically the
differential cross section for inclusive QE $(e,e')$ processes is
written in the form \be
\left[\frac{d\sigma}{d\epsilon'd\Omega'}\right]_{(e,e')}\cong
\overline{\sigma}^{e}(q,\omega;p=|y|,{\cal E}=0)\cdot F(q,\omega) \,
, \label{scaling} \ee where the single-nucleon cross section is
evaluated at the special kinematics $p=|y|$ (with $y$ the scaling
variable; see the next section) and ${\cal E}=0$ (the residual
nucleus in its ground state). This corresponds to the lowest value
of the missing momentum occurring when ${\cal E}=0$. The term
$\overline{\sigma}^{e}$ refers to the azimuthal-angle-averaged
single-nucleon cross section and it also incorporates the
kinematical factor $K$ in Eq.~(\ref{PWIA}) and the contribution of
all nucleons in the target, {\it i.e.,} $\overline{\sigma}^e\equiv
K\sum_{i=1}^A\int d\phi_{N_i}\sigma^{eN_i}/2\pi$.

The function $F(q,\omega)$ in Eq.~(\ref{scaling}) is known as the
scaling function and it is given in PWIA in terms of the spectral
function: \be F(q,\omega)=
2\pi\int\!\!\!\int_{\Sigma(q,\omega)}p\,dp\, d{\cal E}\,S(p,{\cal
E}) \, . \label{scaling_function} \ee

A detailed study of the scaling function and its connection with the
momentum distribution will be presented in next section. However,
let us start by pointing out some general interesting features of
this basic result. First, only in the case in which it would be
possible to extend the kinematically allowed region
$\Sigma(q,\omega)$ to infinity in the excitation energy plane, {\it
i.e.,} ${\cal E}_{max}\rightarrow \infty$, would the scaling
function be directly linked to the true momentum distribution of the
$A$-nuclear system: \be n(p)\equiv \int_0^\infty d{\cal E} S(p,{\cal
E}).\ee Second, guided by the PWIA result in Eq.~(\ref{scaling}), an
experimental scaling function can be also defined by dividing the
experimental QE $(e,e')$ cross section by the single-nucleon
function, $\overline{\sigma}^e$. At high enough values of the
momentum transfer $q$, the function $F_{exp}(q,\omega)$ has been
shown to satisfy scaling in the region below the QE peak, that is,
$F_{exp}$ becomes only a function of the scaling variable $y$
(see~\cite{Day:1990mf,DS199,DS299,MDS02} for details). Note that
Eq.~(\ref{scaling_function}) does not apply to $F_{exp}(q,\omega)$
which incorporates ingredients not included in the simple PWIA
approach: FSI, MEC, re-scattering processes, {\it etc.} The
contribution of these effects and their impact on the scaling
phenomenon depend on the kinematical region explored, leading in
particular to a significant scaling breaking in the region above the
QE peak.

Furthermore, based on the analysis performed with the Relativistic
Fermi Gas (RFG) model, and making use of the separate longitudinal
($L$) and transverse ($T$) $(e,e')$ data, experimental superscaling
functions have been introduced: \ba f_{exp}(q,\omega) &\equiv& k_F
F_{exp}(q,\omega)
\label{littlef1} \\
f_{exp}^{L(T)}(q,\omega) &\equiv& k_F F_{exp}^{L(T)}(q,\omega)
\label{littlef2} \, , \ea where $k_F$ is the Fermi momentum. In
particular, the $L$ response is thought to have very little
contribution from meson production and from meson-exchange currents
and thus should be the place where the underlying nuclear dynamics
can cleanly be resolved. It has been shown to superscale, {\it
i.e.,} the function $f_{exp}^L$ shows only a very mild dependence
upon the momentum transfer $q$ (first-kind scaling) and the nuclear
system considered (second-kind scaling). This has led to introduce a
universal experimental superscaling function that constitutes a
strong constraint for any theoretical model describing QE electron
scattering. Not only should the superscaling behavior be fulfilled,
but also the specific shape of $f_{exp}^L$ must be reproduced. This
subject has been studied in detail in previous work showing the
importance of FSI and
relativity~\cite{neutrino2,PRL05,jac06,Amaro07,jac_plb}, and those
studies clearly show that any conclusion about the momentum
distribution based on Eq.~(\ref{scaling_function}) should be taken
with caution. Being aware of this, it is illustrative, however, to
analyze in detail the basic approaches on which the ``link'' between
the momentum distribution and the scaling (superscaling) function is
based. Moreover, the usual analysis restricted in the past to the
region below the QE peak, is now extended to the region above the
peak, since the superscaling function $f_{exp}^L$ is defined for
both negative and positive values of the scaling variable (see
discussion in next section).

\section{The scaling function}\label{sect2}

As already shown, in PWIA the scaling function can be expressed as
an integral of the spectral function $S$ in the $(p,{\cal E})$ plane
(Eq.~(\ref{scaling_function})), with $p$ the struck nucleon's
momentum and \be {\cal E}(p)\equiv
\sqrt{M_B^{*^2}+p^2}-\sqrt{M_B^{0^2}+p^2} \geq 0 \, , \ee the
excitation energy of the recoiling system $B$, with $M_B^0$ the
ground-state mass of the residual nucleus and $M_B^*$ the general
invariant mass of the daughter final state. The integration in
Eq.~(\ref{scaling_function}) is extended to the kinematically
allowed region in the $(p,{\cal E})$ plane at fixed values of the
momentum and energy transfer, $(q,\omega)$. This is represented by
$\Sigma(q,\omega)$. The general kinematics corresponding to QE
$(e,e')$ processes leads to the following ${\cal E}$-integration
range~\cite{Day:1990mf,scaling88} \be \max\{0,{\cal E}^+\}\leq{\cal
E}\leq{\cal E}^- \, , \ee where \be {\cal
E}^{\pm}(p;q,\omega)=(M_A^0+\omega)-\left[\sqrt{(q\pm p)^2+m_N^2}+
  \sqrt{M_B^{0^2}+p^2}\right]
\ee and where $M_A^0$ is the target nuclear mass and $m_N$ the
nucleon mass.

\begin{figure}[htb]
\begin{center}
\includegraphics[scale=0.8]{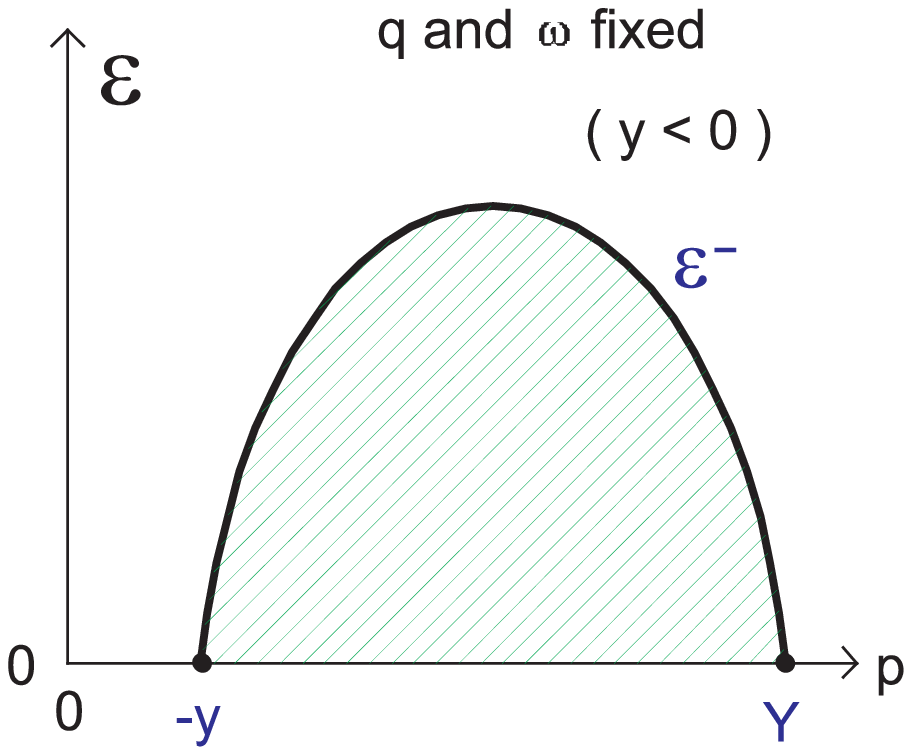}\ \ \ \includegraphics[scale=0.8]{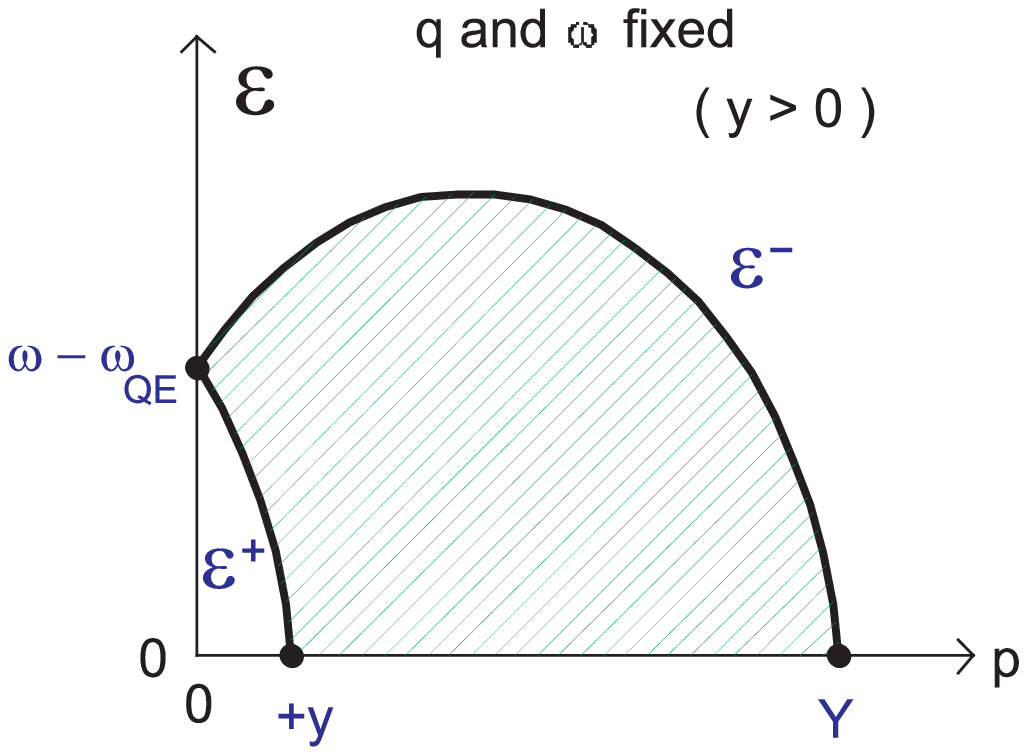}
\caption{(Color online) Excitation energy corresponding to negative
(left) and positive (right) values of $y$.
}
\label{Fig1}
\end{center}
\end{figure}

The intercepts between the curve ${\cal E}^-$ and the $p$-axis will
be denoted by $-y$ and $Y$, {\it i.e.,} ${\cal
E}^-(-y;q,\omega)={\cal E}^-(Y;q,\omega)=0$. The integration region
$\Sigma(q,\omega)$  is shown in Fig.~\ref{Fig1} for fixed values of
the transferred energy and momentum for $\omega<\omega_{QE}$
(left-hand panel) and $\omega>\omega_{QE}$ (right-hand panel), with
$\omega_{QE}$ the energy where the quasielastic peak (QEP) occurs.
In the region below the QEP, $y$ is negative and $p=-y$ represents
the minimum value for the struck nucleon's momentum. Above the QEP
$y$ is positive and the curve ${\cal E}^+$ cuts the integration
region when $p<y$.

In terms of the independent variables $q$ and $\omega$ the intercepts
$\pm y$ and $Y$ are given by
\ba
y(q,\omega) &=& \left\{(M_A^0+\omega)\sqrt{\Lambda^2-M_B^{0^2}W^2}-q\Lambda\right\}/W^2 \label{ysmall}
\\
Y(q,\omega) &=&
\left\{(M_A^0+\omega)\sqrt{\Lambda^2-M_B^{0^2}W^2}+q\Lambda\right\}/W^2
\label{ylarge} \ea with $W\equiv\sqrt{(M_A^0+\omega)^2-q^2}$ the
center of mass energy and $\Lambda\equiv (M_B^{0^2}-m_N^2+W^2)/2$.
Then the scaling function in Eq.~(\ref{scaling_function}) can be
recast as follows \ba {\dfrac{1}{2\pi}}F(q,y) &\!=\!&
\int_{-y}^{Y(q,y)}\!\!p\,dp\int_0^{{\cal E}^-(p;q,y)}d{\cal E}
S(p,{\cal E}) \ \ \ \ \ \ \ \ \ \ \ \ \ \ \ \ \ \ \ \ \ \ \ \ \ \ \
\ \ \ \ \ \ \ \ \ \ \ \ \ \ \mbox{if}\ y<0 \label{eq:Fneg}
\\
{\dfrac{1}{2\pi}}F(q,y) &\!=\!& \int_{0}^y p\,dp\int_{{\cal E}^+(p;q,y)}^{{\cal E}^-(p;q,y)}
d{\cal E} S(p,{\cal E})\!+\!
 \int_{y}^{Y(q,y)}\!\!p\,dp\int_0^{{\cal E}^-(p;q,y)}\!d{\cal E} S(p,{\cal E})
\ \ \ \mbox{if}\ y>0 \label{eq:Fpos} \ea for negative and positive
values of $y$, respectively. The analysis presented in previous work
has been restricted to the negative-$y$ region, {\it i.e.,} below
the QEP, since this is the region where cross section data fulfill
$y$-scaling properties. The function $F_{exp}$ does not scale for
positive values of $y$ because of the significant scaling violations
introduced by effects beyond the IA, namely, inelastic processes and
contributions from meson-exchange current. However, these
contributions mostly reside in the purely transverse response and
are negligible in the $L$ channel. The ``universal'' superscaling
function extracted from the analysis of the separated $L$ data, and
defined for both negative and positive values of the scaling
variable, explains our interest in extending the study to the region
above the QEP. This strategy, which forces us to employ the
superscaling function $f_{exp}^L$ to determine $F_{exp}^L=f_{exp}^L
/k_F$ instead of the usual $y$-scaling function $F_{exp}$, can lead
to significant effects concerning the momentum and energy
distribution in the spectral function, as discussed below.

In the above expressions we have chosen $(p,{\cal E};q,y)$
as independent variables. In terms
of these we can also express the energy transfer \be \omega(q,y) =
\sqrt{(q+y)^2+m_N^2}+\sqrt{M_B^{0^2}+y^2}-M_A^0 \, , \ee the limits
of the excitation energy \be {\cal E}^\pm (p;q,y) =
\left[\sqrt{(q+y)^2+m_N^2}-\sqrt{(q\pm p)^2+m_N^2} \right]
 + \left[\sqrt{M_B^{0^2}+y^2}-\sqrt{M_B^{0^2}+p^2}\right] \label{exc1}
\ee
and the upper limit of $p$:
\be
Y(q,y) = \frac{
M_B^{0^2}(2q+y)
+2(q+y)\sqrt{M_B^{0^2}+y^2}\sqrt{(q+y)^2+m_N^2}
+y\left[2(q+y)^2+m_N^2\right]
}
{M_B^{0^2}+2\sqrt{M_B^{0^2}+y^2}\sqrt{(q+y)^2+m_N^2}
+2y(q+y)+m_N^2
} \, . \label{Ylarge_app}
\ee

In the thermodynamic limit $M_B^0\to\infty$ we get
\ba
{\cal E}^\pm (p;q,y) &\to& \sqrt{(q+y)^2+m_N^2}-\sqrt{(q\pm p)^2+m_N^2}
\equiv E_{q+y}-E_{q\pm p}
\label{eq:Elim}
\\
Y(q,y) &\to& 2q+y \, ,
\label{eq:Ylim}
\ea
where we have introduced the nucleon energies $E_k\equiv\sqrt{k^2+m_N^2}$.
Moreover, notice that in the limit of very large momentum transfer,
{\it i.e.,} $q\gg|y|$ and $q\gg m_N$, the above limiting values reduce to
$Y\to 2q$ and ${\cal E}^\pm \to y\mp p$.

Following previous arguments presented in~\cite{Day:1990mf,Ciofi91},
it is instructive to split the spectral function into two terms,
corresponding to zero and finite excitation energy, respectively:
\be \label{eq:spectral} S(p,{\cal E}) = n_0(p)\delta({\cal E}) +
S_1(p,{\cal E}) \ee with $S_1(p,{\cal E}=0)=0$, which, inserted in
Eqs.~(\ref{eq:Fneg},\ref{eq:Fpos}) yields \ba
{\dfrac{1}{2\pi}}F(q,y<0) &=& \int_{-y}^{Y(q,y)}p\,dp\, n_0(p)+
\int_{-y}^{Y(q,y)}p\,dp\int_0^{{\cal E}^-(p;q,y)}d{\cal E}
S_1(p,{\cal E})
\\
{\dfrac{1}{2\pi}}F(q,y>0) &=&
\int_{y}^{Y(q,y)}p\,dp\, n_0(p) \nonumber \\
&+&\left[
\int_{0}^y p\,dp\int_{{\cal E}^+(p;q,y)}^{{\cal E}^-(p;q,y)}d{\cal E}
+ \int_{y}^{Y(q,y)}p\,dp\int_0^{{\cal E}^-(p;q,y)}d{\cal E}
\right] S_1(p,{\cal E}) \, .
\nonumber\\
\ea

In order to analyze how the scaling function and the nucleon
momentum distribution are connected, we proceed by evaluating the
derivatives of the scaling function $F$ with respect to $y$ and $q$.
Making use of the Leibniz's formula and choosing $(p;q,y)$ as the
three remaining independent variables, after some algebra we finally
get the following results:

\subsection{\bf Negative-$y$ region}
\ba
{\dfrac{1}{2\pi}}\frac{\partial F}{\partial y} &=& Y\,n_0(Y)\left(\frac{\partial Y}{\partial y}\right)
-y\,n_0(-y)+
\int_{-y}^Y p\,dp\,\left(\frac{\partial{\cal E}^-}{\partial y}\right)
S_1(p,\,{\cal E}^-)
\label{eq:dFdyneg}\\
{\dfrac{1}{2\pi}}\frac{\partial F}{\partial q} &=& Yn_0(Y)\left(\frac{\partial Y}{\partial q}\right)
+\int_{-y}^Y p\,dp\,\left(\frac{\partial {\cal E}^-}{\partial q}\right)
S_1(p,\,{\cal E}^-) \, .
\label{eq:dFdqneg}
\ea

Making use of the limits in Eq.~(\ref{exc1}) and assuming the
residual mass $M_B^0$ to be much larger than the momenta, $|y|,p,q$,
we simply have \be \frac{\partial{\cal E}^-}{\partial y} \simeq
\frac{q+y}{E_{q+y}}\,\,, \,\,\,\,\,\,\,\, \frac{\partial{\cal
E}^-}{\partial q}\simeq \frac{q+y}{E_{q+y}}- \frac{q-p}{E_{q-p}} \,
. \ee Likewise, the derivatives of $Y$ reduce to $\partial
Y/\partial y \simeq 1$ and $\partial Y/\partial q \simeq 2$.

Introducing these results in the general expressions in
Eqs.~(\ref{eq:dFdyneg},\ref{eq:dFdqneg}), we get \ba
{\dfrac{1}{2\pi}}\frac{\partial F}{\partial y} &=& Y\,n_0(Y)
-y\,n_0(-y)+ \frac{q+y}{E_{q+y}} \int_{-y}^Y p\,dp\,S_1(p,\,{\cal
E}^-)
\label{eq:dFdyneg1}\\
{\dfrac{1}{2\pi}}\frac{\partial F}{\partial q} &=& 2Yn_0(Y)+
\int_{-y}^Y p\,dp\, \left[ \frac{q+y}{E_{q+y}}-\frac{q-p}{E_{q-p}}
\right] S_1(p,\,{\cal E}^-) \label{eq:dFdqneg1} \ea with ${\cal
E}^-$ and $Y$ given in the thermodynamic limit by
Eqs.~(\ref{eq:Elim}) and (\ref{eq:Ylim}). Note that the
excited-state contribution in the spectral function, that is $S_1$,
is evaluated at energies along the curve ${\cal E}^-$.

\begin{figure}[htb]
\begin{center}
\includegraphics[angle=270,scale=0.5]{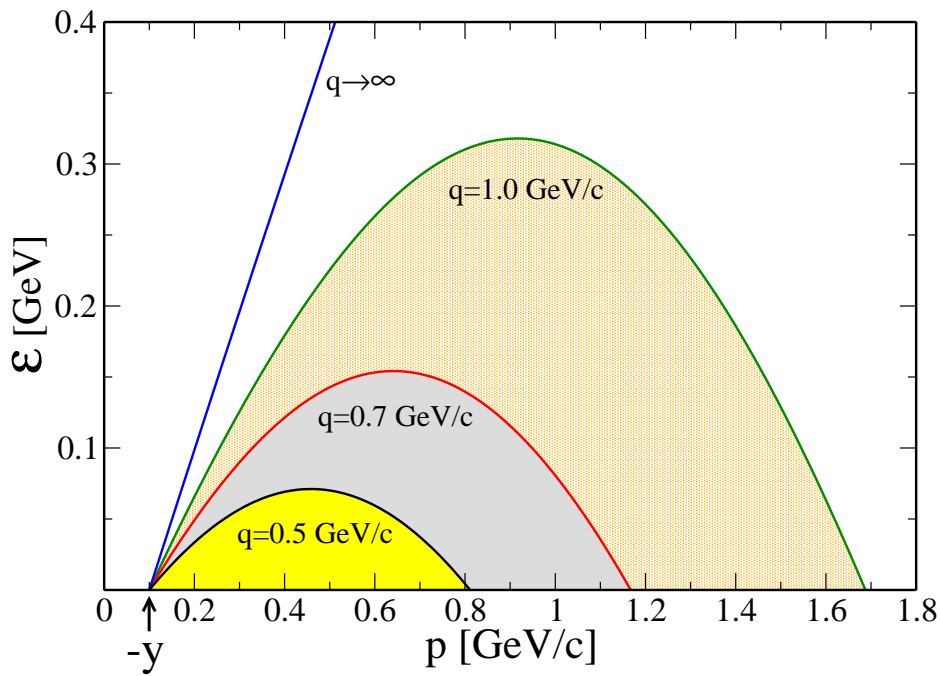}
\caption{(Color online) Integration region in the $({\cal E},p)$
plane for $y=-0.1$ GeV/c and $^{12}$C as the target selected. Each
curve corresponds to ${\cal E}^-$ for a different momentum transfer.
} \label{Fig2}
\end{center}
\end{figure}

\begin{figure}[htb]
\begin{center}
\includegraphics[angle=270,scale=0.5]{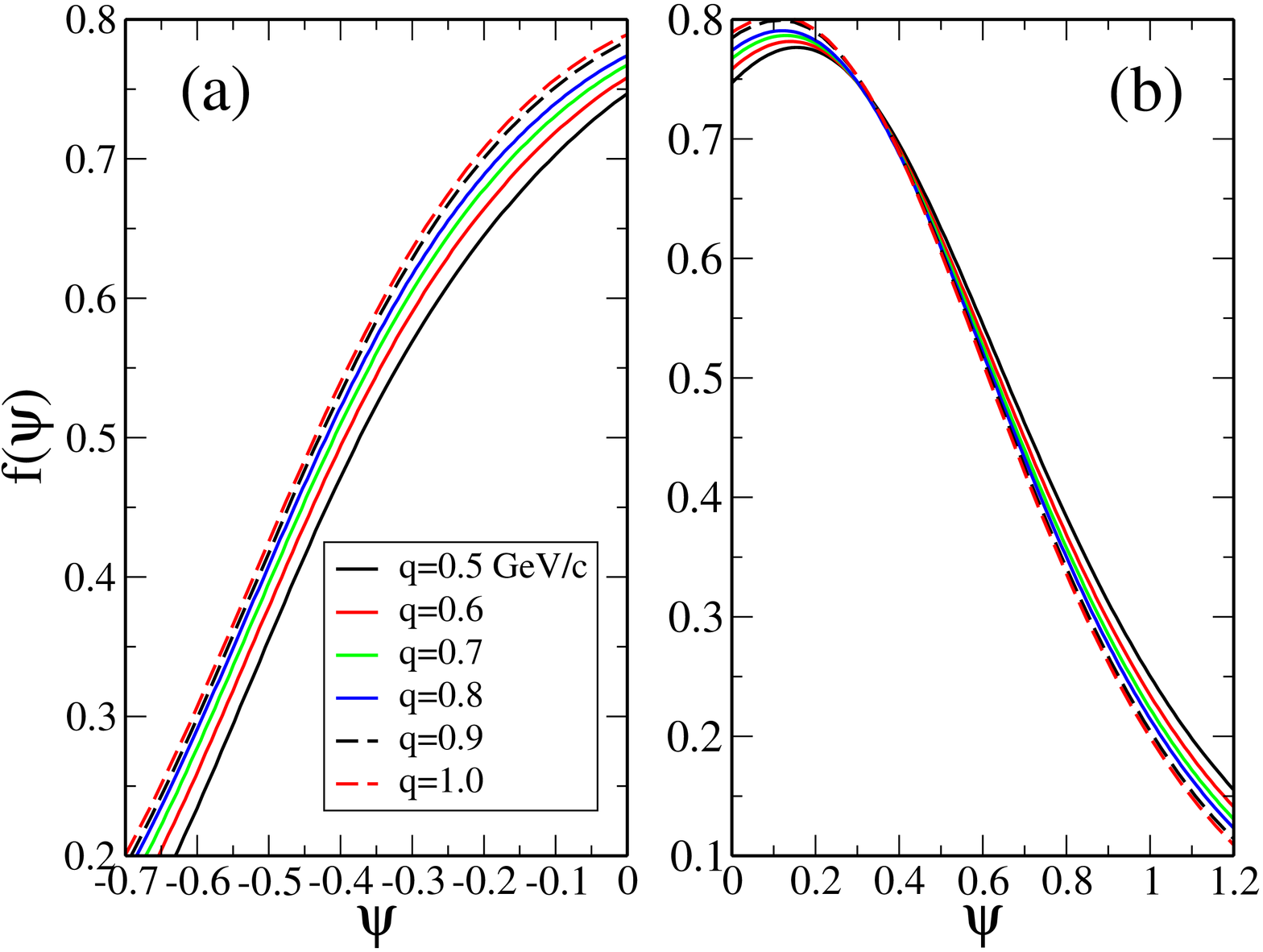}
\caption{(Color online) Superscaling function $f(\psi)$ for negative
(a) and positive (b) values of the scaling variable $\psi$.
Results correspond to $^{12}$C$(e,e')$ evaluated in RPWIA for
different momentum transfers.} \label{Fig3a}
\end{center}
\end{figure}

\begin{figure}[htb]
\begin{center}
\includegraphics[angle=270,scale=0.5]{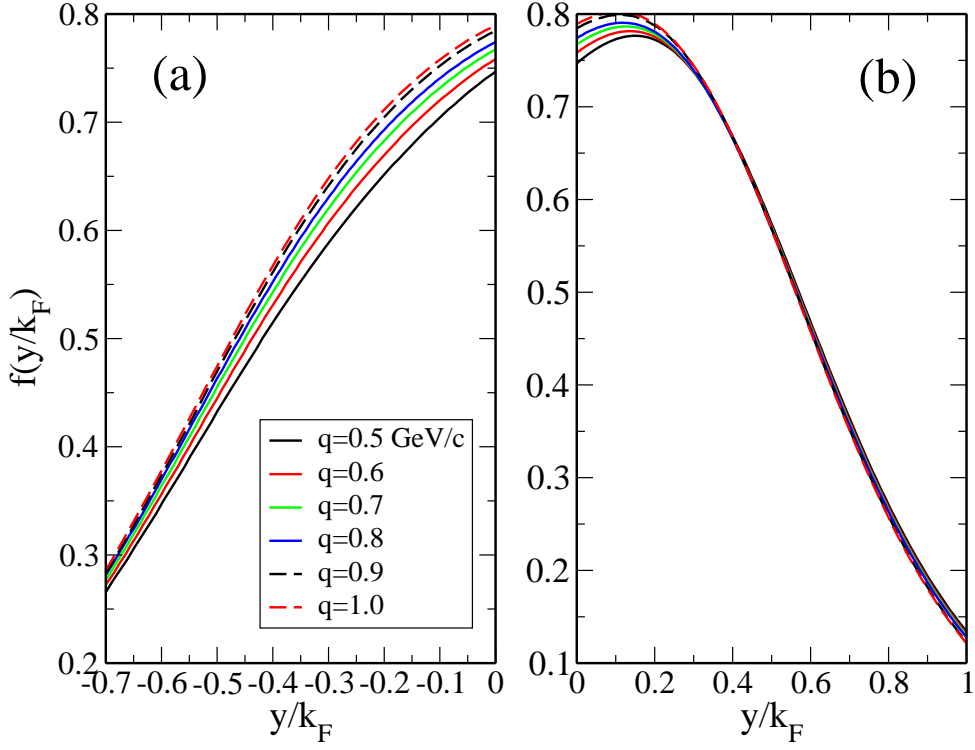}
\caption{(Color online) Superscaling function for negative (a)
and positive (b) values of the dimensionless scaling variable
$y/k_F$.
Results correspond to $^{12}$C$(e,e')$ evaluated in RPWIA for
different momentum transfers.} \label{Fig3b}
\end{center}
\end{figure}

For $q$ sufficiently large, $q\gg -y$, the upper limit $Y$ can be
safely taken to $\infty$ and, since $\lim_{Y\to\infty} Y n_0(Y)=0$,
the expressions for the derivatives simplify to \ba
{\dfrac{1}{2\pi}}\frac{\partial F}{\partial y} &=& -y\,n_0(-y)+
\frac{q +y}{E_{q+y}}\int_{-y}^\infty p\,dp\,S_1(p,\,{\cal E}^-)
\label{eq:dFdyneginf}
\\
{\dfrac{1}{2\pi}}\frac{\partial F}{\partial q} &=&
\int_{-y}^\infty p\,dp\,
\left(\frac{q+y}{E_{q+y}}-\frac{q-p}{E_{q-p}}\right)
S_1(p,\,{\cal E}^-) \, .
\label{eq:dFdqneginf}
\ea

If we further assume that $S_1$ is small for large values of $p$, so
that the main contribution to the integral Eq.~(\ref{eq:dFdqneginf})
comes from $p\simeq -y$, then we get \be \lim_{q\to\infty}
\frac{\partial F}{\partial q} =0 \,, \ee namely scaling of the first
kind (the scaling function $F$ loses its dependence upon $q$).

We also observe that, since at a fixed value of $y$ the integration
region in Eq.~(\ref{eq:dFdqneg1}) increases with $q$ and the
integrand is a positive function, the asymptotic value $F(y)$ is
reached from below ({\it i.e.,} monotonically increasing as a
function of $q$) in any PWIA approach, in contrast with what
experimental data seem to indicate~\cite{DS199,DS299,MDS02}. This is
clearly illustrated in Fig.~\ref{Fig2} where the integration region
is shown for different values of the momentum transfer at fixed $y$,
and it is also consistent with results shown in the left-hand panels (a)
of Figs.~\ref{Fig3a} and~\ref{Fig3b}. In Fig.~\ref{Fig3a} we present
the superscaling function $f(\psi)$ evaluated within the framework
of the Relativistic Plane-Wave Impulse Approximation (RPWIA)
(see~\cite{PRL05,jac06} for details) for different $q$-values and
plotted against the superscaling variable $\psi$ in the
negative-$\psi$ region (below the QEP). This variable is given by
\cite{scaling88,DS299} \be \psi =\frac{1}{\sqrt{\xi_F}}
\frac{\lambda - \tau}{\sqrt{(1+\lambda) \tau + \kappa \sqrt{\tau
(1+\tau)}}} \, , \label{psi-RFG} \ee where $\lambda \equiv \omega /
2 m_N$, $\kappa \equiv q/ 2 m_N$ and $\tau\equiv |Q^2|/ 4 m_N^2 =
\kappa^2 -\lambda^2$. The scaling variables $y$ and $\psi$ are
closely connected~\cite{DS299}: \be\label{eq:y_psi} \psi = \left(
\frac{y}{k_F} \right) \left[ 1
+\sqrt{1+\frac{m_N^2}{q^2}}\frac{1}{2} \eta_F \left( \frac{y}{k_F}
\right) +{\cal O} [\eta_F^2] \right] \simeq \dfrac{y}{k_F}\,, \ee
where $\eta_F=k_F/m_N$ and, as noted above, the superscaling
function $f$ is connected with $F$ via $f\equiv k_F\times F$ with
$k_F$ the Fermi momentum. The curves in Fig.~\ref{Fig3a} may be
compared with the RPWIA results for the superscaling function, now
for negative and positive values of the dimensionless scaling
variable $y/k_F$ obtained using the quadratic form of
Eq.~(\ref{eq:y_psi}); see Fig.~\ref{Fig3b}. As shown, at fixed
$\psi$ (or $y/k_F$) the function $f(\psi)$ increases with $q$ in
accordance with the previous discussion. The basic results shown in
Figs.~\ref{Fig3a} and \ref{Fig3b} demonstrate that $\psi$ and
$y/k_F$ can be used interchangeably as long as one does not focus on
the few percent differences seen in the figures, namely, for large
magnitudes of the scaling variables. 

In showing the results we choose $^{12}$C as an illustrative 
example. Indeed this nucleus is relevant for many neutrino oscillation
experiments, where superscaling ideas can be used to make reliable 
predictions of neutrino-nucleus cross sections~\cite{Amaro:2004bs}.
Moreover, the analysis of the world data performed in \cite{DS199} 
points to an excellent superscaling in the so-called scaling region 
($\psi<0$) for nuclei with $A\geq 12$. 
Note, however, that even the $^4$He data display a very good superscaling 
behavior for large negative values of the scaling variable
($\psi<-0.2$), while at the quasielastic peak there is a 10\% violation
due to the very different spectral function of the lightest nuclei.

\subsection{\bf Positive-$y$ region}

In this case, as shown in Fig.~\ref{Fig1} (right-hand panel), the
integration region in the $(p,{\cal E})$-plane is limited by the two
curves, ${\cal E}^+$ and ${\cal E}^-$, in the missing momentum
region $[0,y]$. This makes the derivative analysis somewhat more
complicated. Moreover, the experimental data show that scaling
arguments of the first kind do not apply to the function
$F(q,\omega)$ in this region, that is, $F$ does not become a
function only dependent on the scaling variable $y$. On the
contrary, it shows a strong dependence upon the momentum transfer
$q$. As already mentioned, this is due to important contributions
beyond the IA contained in the transverse channel.
Therefore, although the analysis that follows is applied to
$F(q,y)$, it should be clearly stated that only the use of the
``universal'' (namely longitudinal) superscaling function $f_L$, and
in particular, the study of its derivative with respect to the
scaling variable in the positive-$y$ region, can reveal important
effects not accounted for by the results obtained in the
negative-$y$ scaling region.

After some algebra, the derivatives of the scaling function $F(q,y)$ are given by
\ba
{\dfrac{1}{2\pi}}\frac{\partial F}{\partial y} &=& Y\,n_0(Y)
\left(\frac{\partial Y}{\partial y}\right)
-y\,n_0(y) \nonumber \\
&+&
\int_0^{Y(q,y)} p\,dp\,S_1(p,\,{\cal E}^-)
\left(\frac{\partial {\cal E}^-}{\partial y}\right) -
\int_0^y p\,dp\,S_1(p,\,{\cal E}^+)
\left(\frac{\partial {\cal E}^+}{\partial y}\right)
\label{eq:dFdypos}\\
{\dfrac{1}{2\pi}}\frac{\partial F}{\partial q} &=& Yn_0(Y)\left(\frac{\partial Y}{\partial q}\right)
\nonumber \\
&+&\int_0^{Y(q,y)} p\,dp\,S_1(p,\,{\cal E}^-)
\left(\frac{\partial {\cal E}^-}{\partial q}\right) -
\int_0^y p\,dp\,S_1(p,\,{\cal E}^+)\left(\frac{\partial {\cal E}^+}{\partial q}\right)
\, .
\label{eq:dFdqpos}
\ea

As in the previous case, from the general expressions for ${\cal
E}^\pm$ given in Eq.~(\ref{exc1}) and assuming the thermodynamic
limit, we get \be \frac{\partial{\cal E}^\pm}{\partial
y}\simeq\frac{q+y}{E_{q+y}}\,\, , \,\,\,\,\, \frac{\partial{\cal
E}^\pm}{\partial q}\simeq\frac{q+y}{E_{q+y}}- \frac{q\pm p}{E_{q\pm
p}} \, , \ee and the derivatives reduce to \ba
{\dfrac{1}{2\pi}}\frac{\partial F}{\partial y} &=&
Y\,n_0(Y)-y\,n_0(y)+ \frac{q+y}{E_{q+y}} \left[ \int_0^Y
p\,dp\,S_1(p,\,{\cal E}^-) - \int_0^y p\,dp\,S_1(p,\,{\cal E}^+)
\right]
\label{eq:dFdypos1}\\
{\dfrac{1}{2\pi}}\frac{\partial F}{\partial q} &=& 2Yn_0(Y)+\frac{q+y}{E_{q+y}}
\left[
\int_0^Y p\,dp\,S_1(p,\,{\cal E}^-)-
\int_0^y p\,dp\,S_1(p,\,{\cal E}^+)
\right]\nonumber\\
&+&
\int_0^y p\,dp\frac{q+p}{E_{q+p}}S_1(p,\,{\cal E}^+)
-\int_0^Y p\,dp\frac{q-p}{E_{q+p}}S_1(p,\,{\cal E}^-) \, .
\label{eq:dFdqpos1}
\ea

Moreover, in the limit of the momentum transfer large enough, $q \gg y$, so
that the condition
$\lim_{Y\to\infty} Y n_0(Y)=0$ holds, the expressions of the derivatives result
\ba
{\dfrac{1}{2\pi}}\frac{\partial F}{\partial y} &=&-y\,n_0(y)+ \frac{q+y}{E_{q+y}}\left[
\int_0^\infty p\,dp\,S_1(p,\,{\cal E}^-) -
\int_0^y p\,dp\,S_1(p,\,{\cal E}^+) \right]
\label{eq:dFdyposlim}\\
{\dfrac{1}{2\pi}}\frac{\partial F}{\partial q} &=&
\int_0^\infty p\,dp\,
\left(\frac{q+y}{E_{q+y}}-\frac{q-p}{E_{q-p}}\right) S_1(p,\,{\cal E}^-)
- \int_0^y p\,dp\,
\left(\frac{q+y}{E_{q+y}}-\frac{q+p}{E_{q+p}}\right) S_1(p,\,{\cal E}^+)
\, . \nonumber \\
\label{eq:dFdqposlim}
\ea

Notice that in the limit in which $y$ can be neglected compared with
$q$, that is, $\dfrac{q+y}{E_{q+y}}\to \dfrac{q}{E_q}$, the same
comment applies to the ratio $(q+p)/E_{q+p}$ involved in the second
integral in Eq.~(\ref{eq:dFdqposlim}), since $p$ is limited within
the range $[0,y]$. Thus, in such a limiting case 
\be \int_0^y
p\,dp\, \left(\frac{q+y}{E_{q+y}}-\frac{q+p}{E_{q+p}}\right)
S_1(p,\,{\cal E}^+) \simeq 0\ \ \ \ \mbox{for}\ \ \ \ q\gg y \ee 
and only the first integral in Eq.~(\ref{eq:dFdqposlim}) survives.
Furthermore, if the spectral function is such that we can neglect
$p$ as compared with $q$ inside the integral, we again get scaling
of the first kind: $ \lim_{q\to\infty} \frac{\partial F}{\partial q}
= 0$.  This is strictly valid only for very large values of $q$ and
it is entirely based on the approximations leading to the expression
in Eq.~(\ref{scaling_function}) that connects the scaling function
with the spectral function. As can be seen from Fig.~\ref{Fig3b}
(panel (b), positive-$y$ region), the RPWIA scaling function shows
a negligible dependence on the momentum transfer for $0.3\lesssim
y/k_F\lesssim 0.8$ ($q\gg y$), whereas for larger $y/k_F$ 
scaling of the first kind begins to be slightly violated. The experimental
scaling function extracted from the analysis of data at intermediate
$q$-values (less than or of the order of the nucleon mass) shows
very important scaling violations in the region above the QEP
(positive values of $y$).

\begin{figure}[htb]
\begin{center}
\includegraphics[angle=270,scale=0.6]{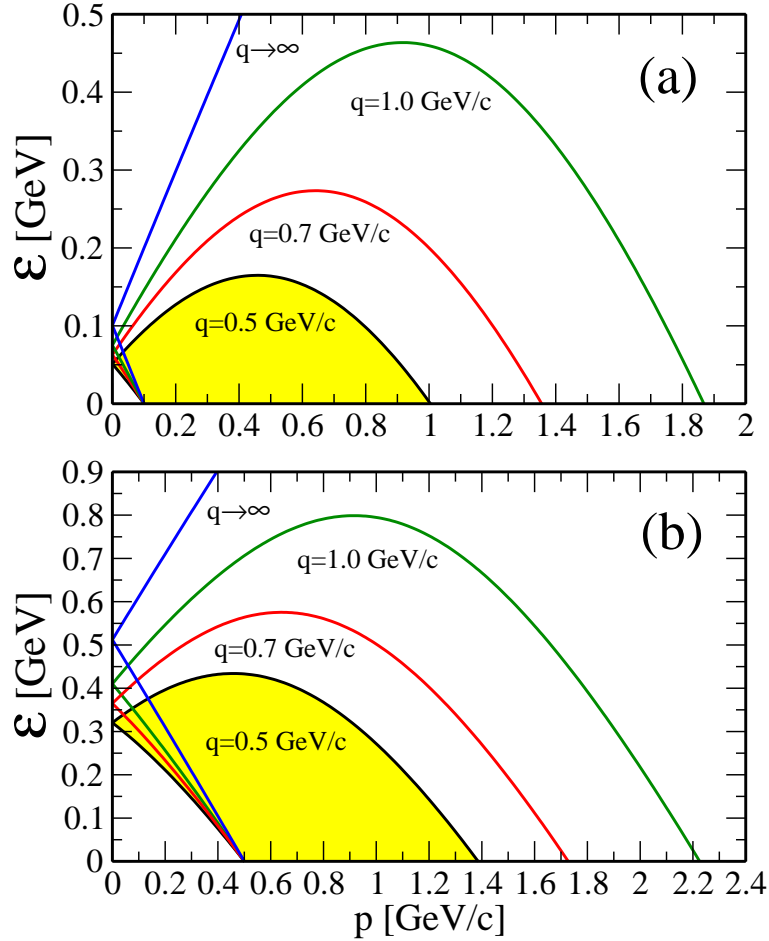}
\caption{(Color online) As for Fig.~\ref{Fig2}, but now for positive
values of $y$. Panel (a) corresponds to $y=0.1$ GeV/c and 
panel (b) to $y=0.5$ GeV/c.} \label{Fig4}
\end{center}
\end{figure}

With regard to the dependence of the scaling function $F$ with $q$
at fixed $y$, we get different behaviors for small and large values
of $y$. Indeed from Eq.~(\ref{eq:dFdqposlim}) we observe that in the
case of $y$ being very small (in the vicinity of zero), the second
integral in Eq.~(\ref{eq:dFdqposlim}) can be neglected. As the
integrand in the remaining integral is positive, we get $\partial
F/\partial q
>0$, {\it i.e.,} the scaling function grows with $q$. This behavior
is in accordance with the one already shown in the negative-$y$
region. On the contrary, for increasing values of $y$ the first
integral in Eq.~(\ref{eq:dFdqposlim}) is expected to diminish
significantly, since the excitation energy curve ${\cal E}^-$ along
which $S_1$ is evaluated lies much higher than ${\cal E}^+$ (see
Fig.~\ref{Fig4}), and it is reasonable to expect that $S_1(p,{\cal
E})$ gets its main contribution for values of the momentum and
energy which are not too large. For $y$ large enough, only the
second integral in Eq.~(\ref{eq:dFdqposlim}) survives, and as its
integrand is also positive, the minus sign in front of it leads to
$\partial F/\partial q <0$, that is, the scaling function $F$
decreases with $q$, changing its behavior with respect to the
previous cases. It is interesting to point out that this result is
consistent with the integration regions shown in Fig.~\ref{Fig4}
where for increasing momentum transfer the curve ${\cal E}^+$ moves
to higher excitation energies in the $({\cal E},p)$ plane. This
means that as $q$ goes up regions at low $({\cal E},p)$ values,
where the spectral function mostly resides, are not kinematically
accessible anymore. A similar argument can be applied to the case of
very small values of $y$ (see panel (a) in Fig.~\ref{Fig4}).
However, here the integration region lost as ${\cal E}^+$ goes up
with increasing $q$ is less important than the effects introduced by
the growing integration region attached to ${\cal E}^-$. This
general behavior is also in accordance with the RPWIA results for
the superscaling function $f$ shown in the right-hand panel of
Fig.~\ref{Fig3b} (positive values of $y$), or, alternatively, the
right-hand panel of Fig.~\ref{Fig3a}. One sees that $f$ increases
with $q$ up to $\psi\gtrsim 0.4$, {\it i.e.,} $y/k_F\sim 0.364$
($q=0.5$~GeV/c), $y/k_F\sim 0.375$ ($q=1.0$~GeV/c), $y/k_F\sim
0.382$ ($q=\infty$~GeV/c) with $k_F=1.2$~fm$^{-1}$ the Fermi
momentum. This corresponds to $y\sim 0.1$~GeV/c, which is the
situation represented in panel (a) of Fig.~\ref{Fig4}. Also note
that the $q$-dependence of $f$ in the region where $y/k_F > 0.3$
seen in Fig.~\ref{Fig3b} is very weak. As observed by examining the
two panels in Fig.~\ref{Fig4}, for large $y$-values the energy
curves ${\cal E}^\pm$ lie very high, and hence, as $q$ increases,
the integrals involved incorporate only additional contributions
which are very small, leading to a very weak variation with momentum
transfer.

\section{Nucleon Momentum Distribution and the Scaling Function}\label{sect3}

In the previous section we have derived general integro-differential
equations connecting the derivatives of the scaling function,
$\partial F/\partial y$ and $\partial F/\partial q$, with the
spectral function. Based on these results applied to both negative
and positive values of $y$, in what follows we revisit the ``usual''
procedure to obtain the nucleon momentum distribution function from
the analysis of QE  $(e,e')$ data. Since the kinematics of electron
scattering lead to
finite integration limits, we may not {\it a priori} draw any strong
conclusions about the ``true'' momentum distribution given as
$n(p)\equiv\int_0^\infty d{\cal E}\,S(p,{\cal E})$, namely the
integral of the spectral function up to infinite excitation energy.
However, assuming the spectral function to reside mostly in the
$(p,{\cal E})$ plane at values of $p$ and ${\cal E}$ which are not
too large, the previous analyses applied to negative- and
positive-$y$ regions lead to different results, thus providing
important and complementary information on how the energy and
momentum are distributed within the spectral function.

The usual procedure considered in previous
work~\cite{Ciofi89,Ciofi91} in order to generate the nuclear
momentum distribution from the scaling function has been based on
the expression: \be n(k)=\left[\frac{-1}{2\pi y}\left(\frac{\partial
F}{\partial y}\right)\right]_{|y|=k} \, , \label{ja1} \ee which has
been widely applied in the negative-$y$ region. In what follows we
extend this study to the positive-$y$ region based on the universal
superscaling function introduced from the analysis of the separated
longitudinal data.

Making use of the general expressions given by
Eqs.~(\ref{eq:dFdyneg1},\ref{eq:dFdypos1})
and assuming the limiting case $\lim_{Y\rightarrow\infty}Yn_0(Y)=0$,
which is valid if the momentum transfer $q$ is sufficiently large,
the momentum distribution functions can be written as follows: \ba
n^{y<0}(q,k) &=& \left [ n_0(-y)- \frac{q+y}{yE_{q+y}}
\int_{-y}^\infty p\, dp\, S_1(p,{\cal E}^{-})\right]_{-y=k}
\nonumber\\
&=&
n_0(k)+ \frac{q-k}{kE_{q-k}} \int_{k}^\infty p\, dp\, S_1(p,{\cal E}^{-})
\label{ja2a}
\\
n^{y>0}(q,k)
&=&
  \left [ n_0(y)- \frac{q+y}{yE_{q+y}}
  \left\{
  \int_{0}^\infty p\, dp\, S_1(p,{\cal E}^{-})-
  \int_0^y p\, dp\, S_1(p,{\cal E}^{+})
  \right\}\right]_{y=k}
\nonumber\\
&=&
  n_0(k)- \frac{q+k}{kE_{q+k}} \left\{
  \int_{0}^\infty p\, dp\, S_1(p,{\cal E}^{-})-
  \int_0^k p\, dp\, S_1(p,{\cal E}^{+})
  \right\}
\, . \label{ja2b}
\ea

As observed, both expressions receive contributions from the $A-1$
system ground state, $n_0(k)$, as well as from the excited states
described through $S_1(p,{\cal E})$. Although using the same
notation for the excitation energy ${\cal E}^-$, note that the
${\cal E}$-curves that enter in the spectral function $S_1$ in
Eqs.~(\ref{ja2a}) and (\ref{ja2b}) are very different (see
Figs.~\ref{Fig2},\ref{Fig4}).

Conclusions on the particular behavior of the previous expressions
can only be drawn based on a specific model for the spectral
function; however, it is illustrative to discuss some general,
``model-independent'', properties. For negative $y$ the function in
Eq.~(\ref{ja2a}) exceeds the purely ground-state contribution, {\it
i.e.,} $n^{y<0}(q,k)>n_0(k)$ for all $q,k$-values. This means that
the contribution from the excited states adds to the ground-state
momentum distribution. Concerning the specific role played by each
one of the two terms in Eq.~(\ref{ja2a}), it is difficult to draw
stringent conclusions without having control over $S_1$. As the
momentum $k$ grows, the contribution of the integral in Eq.~(\ref{ja2a}) is
expected to diminish significantly ($S_1$ mostly residing at momenta
and excitation energies which are not too large).
A similar comment applies also to the ground-state contribution that
decreases as $k$ gets larger. The analysis of Eq.~(\ref{ja2b}) in
the positive-$y$ region differs because of the relative
contributions provided by the two integrals linked to the excited
states. In this case the global response $n^{y>0}(q,k)$ can be
smaller and/or larger than the purely ground-state contribution,
$n_0(k)$, depending on the specific missing momentum value.

In what follows we discuss in detail some particular situations,
thereby drawing some preliminary conclusions on the general behavior
shown by $n^{y \lessgtr 0}(q,k)$. Let us start by considering the
value of the nucleon momentum $k$ to be in the vicinity of zero.
Thus, neglecting $k$ as compared with the momentum transfer $q$
$(k\ll q)$ and assuming $\int_0^\infty p\,dp\,S_1(p,{\cal E}^-)\gg
\int_0^{k}p\,dp\,S_1(p,{\cal E}^+)\to 0$, we can write: \ba
n^{y<0}(q,k)&\simeq &
  n_0(k)+ \frac{q}{kE_q}
    \int_{k}^\infty p\, dp\, S_1(p,{\cal E}^-)
    > n_0(k) \, , \label{negat} \\
n^{y>0}(q,k)&\simeq &
 n_0(k)- \frac{q}{kE_q}
  \int_{0}^\infty p\, dp\, S_1(p,{\cal E}^-)
   <n_0(k) \,. \label{posit}
\ea From these results the following relation (valid for $k$ small
enough) occurs, \be n^{y>0}(q,k)\leq n_0(k)\leq n^{y<0}(q,k) \,
.\label{condition} \ee Moreover, from Eqs.~(\ref{negat},\ref{posit})
the ground-state contribution is roughly given as $n_0(k)\simeq
[n^{y<0}+n^{y>0}]/2$.

As the nucleon momentum $k$ grows, the two functions $n^{y<0}(q,k)$
and $n^{y>0}(q,k)$ in Eqs.~(\ref{ja2a},~\ref{ja2b}) get closer,
crossing each other at some specific $k$, such that
$n^{y>0}(q,k)>n^{y<0}(q,k)$ for larger $k$. From the integration
region in the $({\cal E}-p)$ plane shown in Fig.~\ref{Fig4}, and
assuming most of the strength in the spectral function to be located
at not too high $p$ and ${\cal E}$, we can conclude that for
intermediate-to-high missing momentum values the main contribution
in $n^{y>0}(q,k)$ comes from the second integral in
Eq.~(\ref{ja2b}), that is, $n^{y>0}(q,k)\simeq
\frac{q+k}{kE_{q+k}}\int_0^k p\,dp\,S_1(p,{\cal E}^+)$.

\begin{figure}[htb]
\begin{center}
\includegraphics[angle=270,scale=0.5]{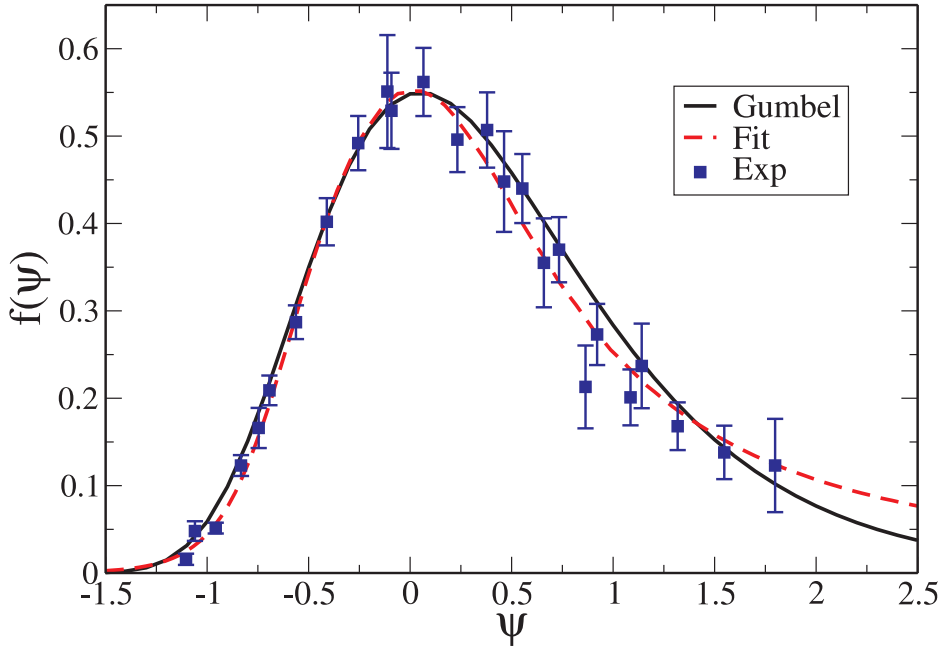}
\caption{(Color online) Average $f_L^{exp}(\psi)$ compared with the
Gumbel distribution  in Eq.~(\ref{s2e17}) (solid) and a
fit of the experimental data (dashed).
}
\label{Fig6}
\end{center}
\end{figure}

\begin{figure}[tb]
\begin{center}
\includegraphics[scale=0.5]{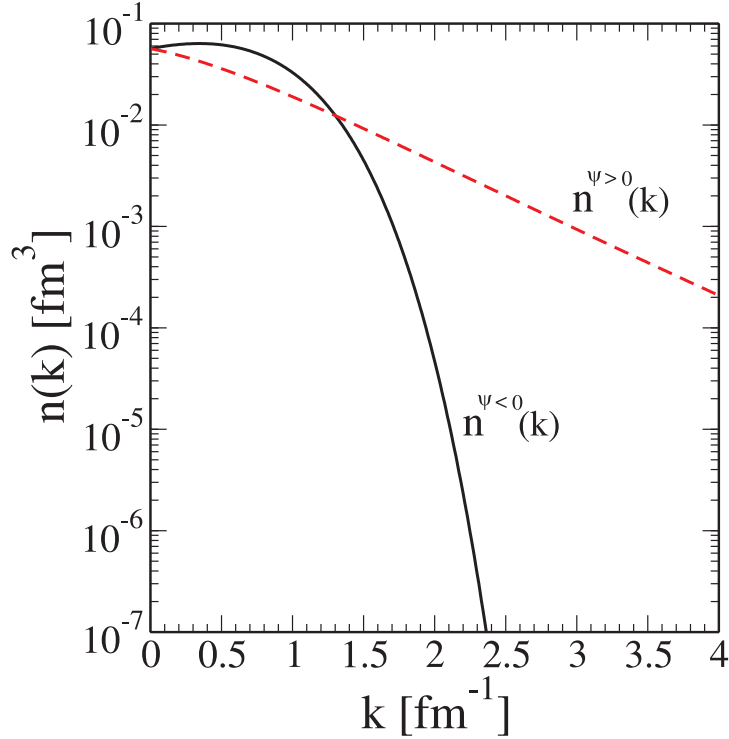}
\caption{(Color online) Nucleon momentum distribution extracted
through the derivative of the superscaling function given by the
Gumbel probability density in Eq.~(\ref{s2e17}). Results
corresponding to negative (solid line) and positive (dashed) values
of the scaling variable are compared. } \label{Fig7}
\end{center}
\end{figure}

\begin{figure}[htb]
\begin{center}
\includegraphics[angle=270,scale=0.5]{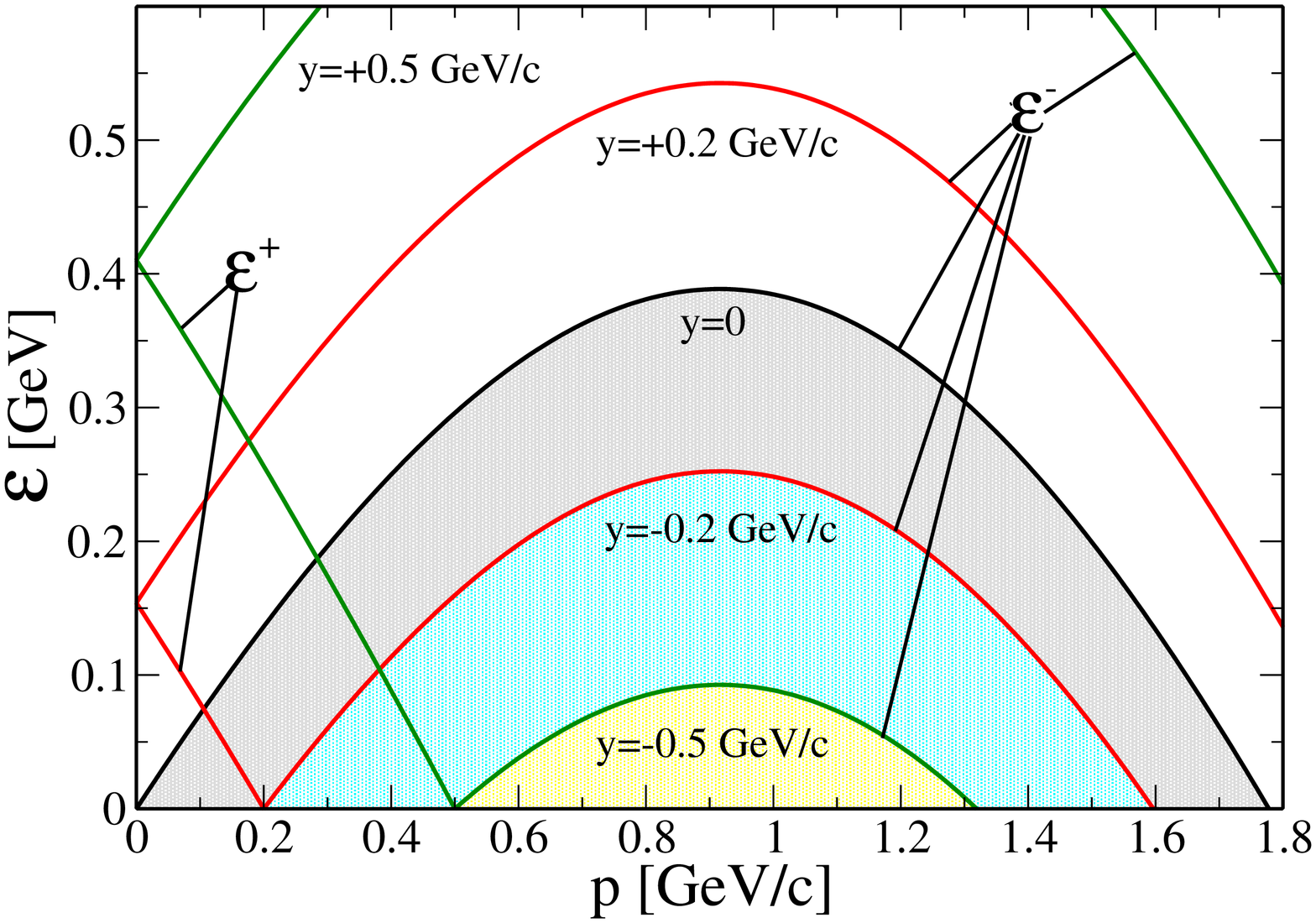}
\caption{(Color online) Integration region in $({\cal E},p)$ plane for $q=1$ GeV/c and different,
negative and positive, values of the scaling variable $y$. The contour curves
${\cal E}^\pm$ in both regions are signaled.
}
\label{Fig8}
\end{center}
\end{figure}

To prove these general properties, in what follows we present
results based on the derivative analysis making use of the
superscaling function $f(\psi)$. In order to simplify the
calculations we represent $f(\psi)$ by means of the Gumbel
probability density function ({\it i.e.,} the derivative of the
Gumbel distribution): 
\begin{equation}\label{s2e17}
f_G(\psi)=\dfrac{1}{\sigma}\exp\left[-\dfrac{(\psi-\mu)}{\sigma}\right]\exp\left[-\exp\left[-\dfrac{(\psi-\mu)}{\sigma}\right]\right].
\end{equation}
In our case the values of the parameters are $\mu=0$ and
${\sigma}=0.67$ {($f_G^{\max}=f_G(0)=0.55$)}. In Fig.~\ref{Fig6} we
compare the Gumbel distribution [Eq.~(\ref{s2e17})] with $f_{exp}^L(\psi)$ and a fit of
the experimental data ~\cite{MDS02}. As shown, the Gumbel
distribution nicely fits the data. Moreover, it fulfills the
unitarity condition $\int_{-\infty}^{+\infty} f(\psi) d\psi=1$. The
nucleon momentum distribution is evaluated through the derivative of
the scaling function
by using Eq.~(\ref{ja1}) and recalling that $f=k_F F$, thus getting
\begin{equation}\label{s2e18}
n(k)=\left[-\dfrac{1}{2\pi y}\dfrac{1}{k_F}
\dfrac{df(\psi(y))}{dy}\right]_{|y|=k},
\end{equation}
that, using the approximate relation $\psi\simeq\dfrac{y}{k_F}$, can
be presented in the form
\begin{equation}\label{s2e19}
n(k)=-\dfrac{1}{2\pi k}\dfrac{1}{k_F}
\left[\dfrac{df(\psi)}{d(k_F|\psi|)}\right]_{k_F|\psi|=k}
\, .
\end{equation}

Note that if the superscaling function is not symmetric with respect to $\psi$,
as is the case for the experimental data, the above expression yields different
momentum distributions for negative and positive values of $\psi$, which will
be denoted by $n^<$ and $n^>$, respectively. On the contrary symmetric scaling
functions, like the RFG one, lead to $n^<=n^>$.

In the case of the Gumbel distribution we get (setting $\mu=0$):
\begin{equation}
\frac{df_G(\psi)}{d\psi} = \frac{1}{\sigma}
\left(e^{-\psi/\sigma}-1\right) f_G(\psi) \, ,
\end{equation}
which leads to
\begin{eqnarray}
n_G^<(k) &=& \frac{1}{2\pi\sigma k_F^2 k} \left[e^{k/(\sigma k_F)}-1\right] f_G(-k/k_F)
\label{nGneg}
\\
n_G^>(k) &=& \frac{1}{2\pi\sigma k_F^2 k} \left[1-e^{-k/(\sigma
k_F)}\right] f_G(k/k_F) \, .\label{nGpos}
\end{eqnarray}

In Fig.~\ref{Fig7} we present the results for
$n^{\psi<0}(k)=\dfrac{n_G^<(k)}{2}$ (solid line) and
$n^{\psi>0}(k)=\dfrac{n_G^>(k)}{2}$ (dashed line) with $n_G^<(k)$
and $n_G^>(k)$ given in Eqs.~(\ref{nGneg},\ref{nGpos}) (at
$k_F=1.2$~fm$^{-1}$).
As expected, $n_G^<(k)$ and $n_G^>(k)$ (and $n^{\psi<0}$ and
$n^{\psi >0}$, respectively) coincide in the limiting case $k=0$:
\begin{equation}
n_G^>(0)=n_G^<(0)=\frac{1}{2\pi\sigma^3 k_F^3 e} \, .
\end{equation}
For missing momenta up to $k\sim 1$ fm$^{-1}$ the main contribution
resides in $n^<$ that is in accordance with Eq.~(\ref{condition})
and the general discussion presented above. At $k\simeq 1.3-1.4$
fm$^{-1}$, {\it i.e.,} $k$ close to the Fermi momentum, $n^<$ and
$n^>$ cross each other, with $n^>$ being much higher for larger
$k$-values.
In fact, whereas $n^<$ shows a steep slope when $k$ increases, which
is in accordance with results based on independent-particle model
descriptions, $n^>$ presents a high momentum tail very far from
$n^<$ and hence from shell-model results (see next section). As
already explained above, this tail at intermediate-to-high $k$ is
linked to the much larger contribution given by the spectral
function $S_1$ when evaluated along the curve ${\cal E}^+$ instead
of ${\cal E}^-$. This general behavior is illustrated in
Fig.~\ref{Fig8} where the contour curves ${\cal E}^\pm$
corresponding to positive- and negative-$y$ values are presented.
The presence of the tail at high momentum
values in the nucleon momentum distribution is a clear signature of
the importance of nucleon-nucleon correlations. Since the spectral
function maps very different regions in the $({\cal E}-k)$ plane for
negative and positive $y$ (Fig.~\ref{Fig8}), the joint analysis of
the two kinematical regions can provide important clues in the
knowledge of NN correlations. 
It should be pointed out that the
functions $n^{\psi<0}(k)$ and $n^{\psi>0}(k)$ evaluated through
Eq.~(\ref{s2e19}) and presented in Fig.~\ref{Fig7} are normalized to
different values connected with the different areas subtended by the
Gumbel distribution function $f_G(\psi)$ at negative and positive
$\psi$, {\it i.e.,} 0.37 (for $\psi<0$) and 0.63 ($\psi>0$).

In particular, it has been shown in \cite{PRL05,Amaro07} in the framework
of relativistic nuclear models that the large positive-$\psi$ tail
of the scaling function is closely related to final-state interactions,
while the negative-$\psi$ region is more affected by initial-state 
correlations, as will be also shown in the next Section in the CDFM model. 
The possibility of connecting different aspects of the momentum distribution 
to initial- and final-state physics will be further explored in future work.

\section[]{Nucleon momentum distribution within the Coherent Density
Fluctuation Model\label{CDFM_results}}

In this section we give, as an example, the results for the nucleon
momentum distribution extracted from the scaling function, obtained
within the framework of a particular nuclear model, namely the
Coherent Density Fluctuation Model (CDFM)~\cite{ant01,ant02}. The
latter is a natural extension to finite nuclei of the relativistic
Fermi gas (RFG) model within which the scaling variables $\psi'$ was
introduced\footnote{The scaling variable $\psi'$ differs from $\psi$
by a phenomenological energy shift $E_s\simeq 20$ MeV (for $^{12}$C)
introduced in order to reproduce the experimental position of the
quasielastic peak: $\psi'(q,\omega) = \psi(q,\omega-E_s)$.}. The
CDFM is based on the generator coordinate method~\cite{ant03} and
includes long-range NN correlations (LRC) of collective type.
In~\cite{ant04,ant05} the scaling function was defined within the
CDFM using the RFG scaling
function~\cite{scaling88,ant11,ant12,ant13} and applied it to
various processes~\cite{ant04,ant05,ant06,ant07,ant08,ant09}.

In the CDFM model~\cite{ant01,ant02}, the one-body density matrix
$\rho (\mathbf{r},\mathbf{r'})$ is an infinite superposition of
one-body density matrices $\rho_x(\mathbf{r},\mathbf{r'})$
corresponding to single Slater determinant wave functions of systems
of $A$ free nucleons homogeneously distributed in a sphere with
radius $x$, density $\rho_0(x)\equiv \dfrac{3A}{4\pi x^3}$, and
Fermi momentum $k_F(x)\equiv
\left[\dfrac{3\pi^2}{2}\rho_0(x)\right]^{1/3}\equiv\dfrac{\alpha}{x}$
(with
$\alpha\equiv\left(\dfrac{9\pi}{8}A\right)^{1/3}\cong1.52A^{1/3}$):
\begin{equation}\label{s2e01}
\rho({\mathbf{r}},{\mathbf{r'}})=\int\limits_0^\infty
|F(x)|^2\rho_x({\mathbf{r}},{\mathbf{r'}}) dx.
\end{equation}
The weight function $|F(x)|^2$ can be expressed in an equivalent way
either by means of the density
distribution~\cite{ant01,ant02,ant05},
\begin{equation}\label{s2e02}
|{F}(x)|^{2}=-\frac{1}{\rho_0(x)} \left. \frac{d\rho(r)}{dr}\right |_{r=x}\mbox{ at }\frac{d\rho(r)}{dr}\leq0,
\end{equation}
or by the nucleon momentum distribution~\cite{ant05},
\begin{equation}\label{s2e03}
|{F}(x)|^{2}=-\frac{3\pi^{2}}{2}\frac{\alpha}{x^{5}} \left.
\frac{dn(k)}{dk}\right |_{k={\alpha}/{x}}\mbox{ at
}\frac{dn(k)}{dk}\leq0.
\end{equation}
In Eqs.~(\ref{s2e02}) and~(\ref{s2e03})
\begin{gather}
\int\rho(\mathbf{r})d\mathbf{r}=A,\quad\int
n(\mathbf{k})d\mathbf{k}=A,~\text{and}\notag\\
\int\limits_{0}^{\infty}|F(x)|^{2}dx=1. \label{s2e04}
\end{gather}
In the version of the CDFM approach suggested in~\cite{ant04,ant05},
the scaling function has the form
\begin{equation}\label{s2e05}
f(\psi')= \int\limits_{0}^{\alpha/(k_{F}|\psi'|)} |F(x)|^{2}
f_{RFG}(x,\psi')dx,
\end{equation}
where the RFG scaling function is
\begin{align}
f_\text{RFG}(x,\psi') =& \displaystyle \frac{3}{4} \left[\! 1\!-\!\left(\!
\frac{k_Fx|\psi'|}{\alpha}\! \right)^{2}\!\right]\! \left\{\! 1\!+\! \left(
\!\frac{xm_N}{\alpha}\!\right)^2 \!\left(\! \frac{k_Fx|\psi'|}{\alpha}
\!\right)^2 \right. \nonumber\\
& \times \displaystyle \left. \left[2+ \left( \frac{\alpha}{xm_N}
\right)^2- 2\sqrt{1+ \left( \frac{\alpha}{xm_N} \right)^2}\right]
\right\}. \label{s2e06}
\end{align}

In the CDFM the Fermi momentum $k_F$ is calculated for each nucleus by
\begin{equation} \label{s2e07}
k_F= \int_{0}^{\infty}  k_{F}(x)|F(x)|^2dx=
\int_{0}^{\infty}  \frac{\alpha}{x}|F(x)|^{2}dx
\end{equation}
and is not a fitting parameter, as it is in the RFG model.

By using Eqs.~(\ref{s2e02}) and~(\ref{s2e03}) in Eqs.~(\ref{s2e05}) and~(\ref{s2e07}),
the CDFM scaling function $f(\psi')$ and $k_F$ can be expressed equivalently by the
density and momentum distributions~\cite{ant05}:
\begin{equation}\label{s2e08}
f(\psi^{\prime})=\frac{4\pi}{A}\int\limits_{0}^{\alpha/(k_{F}|\psi^{\prime}|)}
\rho(x)\left
[x^{2}f_\text{RFG}(\psi^{\prime},x)+\frac{x^{3}}{3}\frac{df_\text{RFG}(\psi^{\prime},x)}{dx}\right
]dx,
\end{equation}
where $f_\text{RFG}(\psi^{\prime},x)$ is given by Eq.~(\ref{s2e06}), and
\begin{gather}
f(\psi^{\prime})=\frac{4\pi}{A}\int\limits_{k_{F}|\psi^{\prime}|}^{\infty}
n(k)\left
[k^{2}f_\text{RFG}(\psi^{\prime},k)+\frac{k^{3}}{3}\frac{df_\text{RFG}(\psi^{\prime},k)}{dk}\right
],\label{s2e09}
\end{gather}
where
\begin{multline}
    f_\text{RFG}(\psi^{\prime},k)=\frac{3}{4} \left [ 1-\left
(\frac{k_{F}|\psi^{\prime}|}{k}\right )^{2}\right ]\left \{1+\left
(\frac{m_{N}}{k}\right )^{2}\left
(\frac{k_{F}|\psi^{\prime}|}{k}\right )^{2}\right.
\times\\\times\left.\left [2+ \left (\frac{k}{m_{N}}\right
)^{2}-2\sqrt{1+\left (\frac{k}{m_{N}}\right )^{2}}\right ]\right
\}.
\label{s2e10}
\end{multline}
Eq.~(\ref{s2e09}) is valid under the condition
\begin{gather}\label{s2e11}
\lim_{k\rightarrow\infty}\left[n(k)k^3\right]=0.
\end{gather}

>From Eq.~(\ref{s2e09}) one can estimate the possibility to obtain information about
the nucleon momentum distribution from the empirical data for the scaling function.
If we keep only the main term of the RFG scaling function from Eq.~(\ref{s2e10}):
\begin{gather}
f_\text{RFG}(\psi^{\prime},k)\simeq\frac{3}{4} \left [ 1-\left
(\frac{k_{F}\psi^{\prime}}{k}\right )^{2}\right ]\label{s2e12}
\end{gather}
and its derivative
\begin{equation}\label{s2e13}
\frac{\partial f_\text{RFG}(\psi^{\prime},k)}{\partial
{k}}\simeq \frac{3}{2} \frac{\left( k_F |\psi^{\prime}|
\right)^2}{{{k}}^3} ,
\end{equation}
then
\begin{equation}\label{s2e14}
f(\psi^{\prime})\simeq 3\pi \int_{k_F|\psi^{\prime}|}^{\infty}
n(k) {k}^2 \left[ 1 -
\frac{1}{3}\frac{\left( k_F |\psi^{\prime}| \right)^2}{{k}^2}
\right]dk.
\end{equation}
In Eq.~(\ref{s2e14})
\begin{equation}\label{s2e15}
\int n(\mathbf{k}) d\mathbf{k}=1.
\end{equation}

Using Eq.~(\ref{s2e14}), $n(k)$ can be found by solving the integral-differential equation:
\begin{equation}\label{s2e16}
n(k)= -\dfrac{1}{2\pi k^2}\left.\dfrac{\partial f(\psi^\prime)}{\partial (k_F |\psi^\prime|)}\right|_{k_F |\psi^\prime|=k}
-\dfrac{1}{k} \int\limits_{k}^{\infty}dk^\prime n(k^\prime).
\end{equation}

\begin{figure}[t]
\centering
\includegraphics[width=80mm]{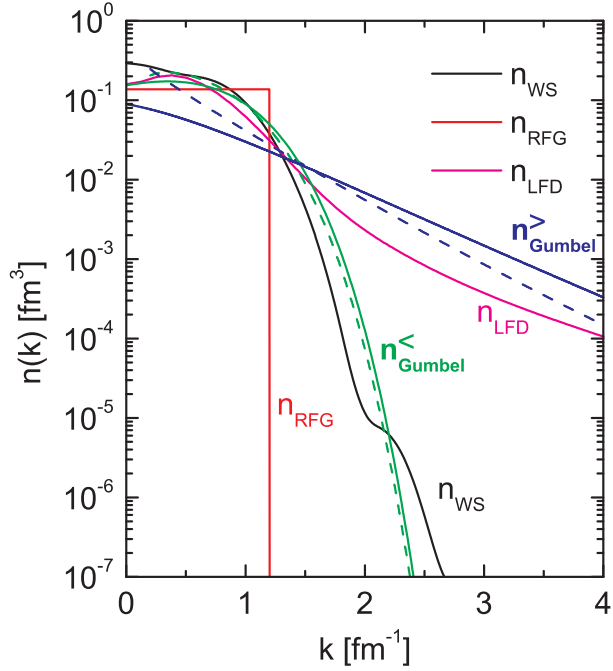}
\caption[]{(Color online) The nucleon momentum distribution
extracted from the scaling function. Solid lines: $n^<$ (light
lines, green online) and $n^>$ (dark lines, blue online) obtained
through the derivative of the scaling function [Eq.~(\ref{s2e19})];
dashed lines: $n^<$ (light lines, green online) and $n^>$ (dark
lines, blue online) using the CDFM integral-differential equation
[Eq.~(\ref{s2e16})]. The Gumbel probability density function
$f_G(\psi)$ [Eq.~(\ref{s2e17})] is used in the calculations. For
comparison are given the momentum distributions from the
Relativistic Fermi Gas model ($n_\text{RFG}$), from the shell model
($n_\text{WS}$) and from the Light Front Dynamics ($n_\text{LFD}$).
All momentum distributions are normalized to unity
[Eq.~(\ref{s2e15})].\label{fig_cdfm}}
\end{figure}

In this work we solve the above equation from CDFM using the
experimentally obtained scaling function. The latter can be
represented the Gumbel probability density function in
Eq.~(\ref{s2e17}). The results for the nucleon momentum distribution
obtained in this way are given in Fig.~\ref{fig_cdfm} by dashed
lines in both cases: $n^<(k)$ for $\psi<0$ (green dashed line) and
$n^>(k)$ for $\psi>0$ (blue dashed line). They are compared with the
results obtained using the expression for $n(k)$ through the
derivative of the scaling function, Eq.~(\ref{s2e19}).

The momentum distributions $n^<(k)$ and $n^>(k)$ obtained by using
Eq.~(\ref{s2e19}) and the experimental scaling function presented by
Eq.~(\ref{s2e17}) are given in Fig.~\ref{fig_cdfm} by solid lines.
For a comparison we present in the same figure the momentum
distributions from the RFG model ($n_\text{RFG}$), the shell-model
results (using Woods-Saxon single-particle wave functions) for
$^{56}$Fe ($n_\text{WS}$), as well as the momentum distribution
($n_\text{LFD}$) obtained within the Light-Front Dynamics (LFD)
approach~\cite{ant14} (see also~\cite{ant05} and the late
modification of the approach in~\cite{ant07}). The latter is based
on the nucleon momentum distribution in the deuteron (including its
high-momentum component) from the LFD method ({\it
e.g.,}~\cite{ant15, ant16} and references therein). In the
calculations $k_F=1.2$~fm$^{-1}$. In Fig.~\ref{fig_cdfm} all nucleon
momentum distributions are normalized to unity [Eq.~(\ref{s2e15})].

One can see from Fig.~\ref{fig_cdfm} that, in general, the results
for $n(k)$ in CDFM confirm the considerations made in
Sects.~\ref{sect1}--\ref{sect3}. Namely,
\begin{itemize}
  \item [(i)] At $k\lesssim 1.3$~fm$^{-1}$ the CDFM momentum distributions
(from Eq.~(\ref{s2e16})) $n^<(k)>n^>(k)$, while at $k\gtrsim
1.3$~fm$^{-1}$ $n^<(k)<n^>(k)$. The same is valid for the momentum
distribution obtained using Eq.~(\ref{s2e19}). This is in accord
with the general consideration from Sect.~\ref{sect3}.
  \item [(ii)] The crossing point of the lines showing $n^<(k)$ and $n^>(k)$
from Eq.~(\ref{s2e19}) is at slightly smaller value of $k$ than that for
$n^<(k)$ and $n^>(k)$ obtained from Eq.~(\ref{s2e16}). This follows also from
the comparison of the explicit forms of Eqs.~(\ref{s2e16}) and~(\ref{s2e19}).
  \item [(iii)] $n^<(k)$ from Eq.~(\ref{s2e16}) is close to the result for
$n^<(k)$ from Eq.~(\ref{s2e19}), while the difference between $n^>(k)$ from
Eq.~(\ref{s2e19}) and $n^>(k)$  from Eq.~(\ref{s2e16}) increases with $k$.
At $k=4$~fm$^{-1}$ $n^>(k)$ from Eq.~(\ref{s2e19}) is around twice larger than
$n^>(k)$ from Eq.~(\ref{s2e16}). At the same time for $k\lesssim 1.2$~fm$^{-1}$,
$n^>(k)$ from Eq.~(\ref{s2e16}) is larger than $n^>(k)$ from Eq.~(\ref{s2e19}).
\end{itemize}

\section{Conclusions}

In the present work a study of the scaling function and its
connection with the momentum distribution is presented. As is well
known, a close relationship between the two quantities exists using
the PWIA and under some conditions for the kinematically allowed
region ($\Sigma(q,\omega)$), once one has accounted for the roles of
FSI, MEC, rescattering processes, {\it etc.} Here these restricted
approximations are considered in detail. The ``usual'' analyses
performed in the past to the region below the QE peak is extended to
the region above the peak, since the superscaling function is
defined for both negative and positive values of the scaling
variable. This is justified, since a ``universal'' superscaling
function has been extracted from the analysis of the separated
longitudinal data. The explicit expressions for the derivatives
$\partial F/\partial y$ and $\partial F/\partial q$ for both
negative- and positive-$y$ regions are derived and their dependences
on $q$ and $y$ are analyzed.

The general integro-differential equations connecting the
derivatives $\partial F/\partial y$ and $\partial F/\partial q$ with
the spectral function are derived. The results obtained allow us to
revisit the ``usual'' procedure to obtain the nucleon momentum
distribution from the analyses of the QE scattering data. The
considerations in the present work lead to results that are quite
different from those obtained solely in the negative-$y$ scaling
region and give information about the energy and momentum
distribution in the spectral function. It is shown that the
expressions for the nucleon momentum distributions $n^{y<0}(q,k)$
and $n^{y>0}(q,k)$ have contributions from the momentum distribution
$n_0(k)$ of the ground state of the system with $A-1$ nucleons, as
well as from the part of the spectral function $S_1(p,{\cal E})$
that contains  information about  the excited states. It is shown
that for small momenta $k$: $n^{y>0}(q,k) \leqq n_0(k) \leqq
n^{y<0}(q,k)$, while  as $k$ grows the two functions $n^{y<0}(q,k)$
and $n^{y>0}(q,k)$ get closer, crossing each other at some value of
$k$ and yielding $n^{y>0}(q,k) > n^{y<0}(q,k)$ for higher $k$.

The general properties of the momentum distribution established in
the present work are validated by the results obtained from the
derivative analysis using the superscaling function $f(\psi)$
represented by the parameterized Gumbel probability  density
function that provides a good fit to the experimental longitudinal
scaling function $f_\text{exp}^L(\psi)$. It is concluded that the
high-momentum tail of the momentum distribution is a clear signature
for the important effects stemming from nucleon-nucleon
correlations.

The general properties of the nucleon momentum distribution obtained
are also illustrated using the scaling function obtained in the
framework of a particular nuclear model, namely the Coherent Density
Fluctuation Model (CDFM) that includes collective long-range NN
correlations. It is shown that the momentum distribution in the CDFM
has the properties already pointed out in the general consideration.

\subsection*{Acknowledgements}

This work was partially supported by DGI (MICINN-Spain) contract
FIS2008-04189, PCI2006-A7-0548, the Spanish Consolider-Ingenio
programme CPAN (CSD2007-00042), by the Junta de Andaluc\'{\i}a, and
by the INFN-CICYT collaboration agreements INFN08-20 \&
FPA2008-03770-E/INFN, as well as by the Bulgarian National Science
Fund under contracts nos~DO--02--285 and DID--02/16--17.12.2009.
M.V.I. acknowledges support from the European Operational programm
HRD through contract BGO051PO001/07/3.3-02/53 with the Bulgarian
Ministry of Education. This work is also supported in part (T.W.D.)
by the U.S. Department of Energy under cooperative agreement
DE-FC02-94ER40818.

\end{document}